\documentclass{appolb}
\usepackage{graphicx}
\usepackage{amsmath, amssymb}
\usepackage{physics}
\usepackage{bbold}
\usepackage{xcolor}
\usepackage{authblk}
\usepackage{mathtools}
\usepackage{subcaption}
\usepackage{cite}

\begin{document}
\title{Control of the Scattering Properties of Complex Systems By Means of Tunable Metasurfaces
}

\author[1]{Jared Erb \thanks{Corresponding author: jmerb@umd.edu}}
\author[2]{David Shrekenhamer}
\author[2]{Timothy Sleasman}
\author[3,4]{Thomas M. Antonsen}
\author[1,4]{Steven M. Anlage}

\affil[1]{\footnotesize Quantum Materials Center, Department of Physics, University of Maryland, College Park, Maryland 20742-4111, USA}
\affil[2]{\footnotesize Johns Hopkins University Applied Physics Laboratory, Laurel, MD, 20723, USA}
\affil[3]{\footnotesize Department of Physics, University of Maryland, College Park, Maryland 20742-4111, USA}
\affil[4]{\footnotesize Department of Electrical and Computer Engineering, University of Maryland, College Park, Maryland 20742-3285, USA}

\maketitle
\begin{abstract}
We demonstrate the ability to control the scattering properties of a two-dimensional wave-chaotic microwave billiard through the use of tunable metasurfaces located on the interior walls of the billiard. The complex reflection coefficient of the metasurfaces can be varied by applying a DC voltage bias to varactor diodes on mushroom-shaped resonant patches, and this proves to be very effective at perturbing the eigenmodes of the cavity. Placing multiple metasurfaces inside the cavity allows us to engineer desired scattering conditions, such as coherent perfect absorption (CPA), by actively manipulating the poles and zeros of the scattering matrix through the application of multiple voltage biases.  We demonstrate the ability to create on-demand CPA conditions at a specific frequency, and document the near-null of output power as a function of four independent parameters tuned through the CPA point.  A remarkably low output-to-input power ratio of $\frac{P_{out}}{P_{in}} = 3.71 \times 10^{-8}$ is achieved near the CPA point at 8.54 GHz.
\end{abstract}
  
\section{Introduction}

We consider bounded complex scattering environments, coupled to the outside world through a finite number of scattering channels.  Examples include enclosed three dimensional spaces such as rooms, cabins in a ship or aircraft, or larger enclosed spaces such as warehouses.  Other examples include two-dimensional microwave billiards and one-dimensional cable graphs with multiple propagation paths between any two points in the billiard or nodes of the graph.  The scattering channels can be coupled to the system through antennas, probes, apertures on the walls of the enclosure, or any means by which wave energy can leave the enclosure and propagate outside it. 
We assume that these systems are reverberant in the sense that waves propagate across the length and breadth of the system multiple times before significantly decaying in amplitude.  Such systems are characterized by a scattering ($S$) matrix that relates the set of in-going wave excitations on the channels to the corresponding set of out-going waves on the same channels.  Because the scattering environment is complicated and typically lossy, the $S$-matrix is sub-unitary and has complex matrix elements that are rapid and irregular functions of the frequency of the waves.

The question arises as to how to control, or tame, a complex scattering environment, such that it can be harnessed to perform specific and useful tasks.  Example tasks include establishing and maintaining a robust communication link between two points inside the enclosure, or transferring wave energy to a specific object inside the enclosure with high efficiency and minimal interference.  It is our belief that active and tunable metasurfaces can be used to alter the scattering properties of complex structures and thus create new opportunities to manipulate complex waves.  Active metasurfaces have the property that they can alter the reflection coefficient of one portion of the boundary of a scattering environment \cite{Chen16,Elsway23}.  This establishes a degree of control of the walls of the enclosure, creating the opportunity to alter the waves everywhere in the enclosure, due to its reverberant nature.

The majority of work on active metasurfaces concerns single-pass reflection or transmission interactions between waves and the metasurface, or involve metasurface antennas that launch waves but do not interact with them again \cite{Faenzi19}.  Here we are concerned with the more challenging situation in which the same waves interact with the same active metasurface multiple times during their propagation.  Efforts to control the wave properties of enclosures by means of active and tunable metasurfaces are relatively few in number.  Gros, \textit{et al.}, showed that three programmable metasurfaces on the walls of a regular six-sided enclosure could be used to effectively stir the modes of the cavity to create a set of uncorrelated cavity configurations \cite{Gros20}.   Frazier, \textit{et al}., placed a 240-element tunable binary metasurface inside a 1 $m^3$ reverberant three-dimensional system, but taking up only 1.5 \% of the surface area, and demonstrated the ability to create 'cold spots' (minima in transmission $S_{21}$) between two arbitrary points inside the enclosure \cite{Frazier20}.  That work showed that coherent perfect absorption (CPA) could also be achieved through variation of the metasurface pixel states, as long as the system was already near the CPA condition at baseline \cite{Frazier20}.  Earlier, one-port perfect absorption was demonstrated in a reverberant environment using a programmable metasurface \cite{Imani20}.  Wavefront shaping with a large number of scattering channels, and creation of CPA states has also been demonstrated in three-dimensional enclosures \cite{dH21PRL,dH21CPA}.  The 240-pixel metasurface experiment also utilized a machine learning algorithm that could find the pixel settings on the metasurface required to generate a desired transmission $S_{21}$ spectrum as a function of frequency over a substantial bandwidth \cite{Frazier22}.  Most recently, an elegant formalism has been created to model the effects of tunable metasurfaces in reverberant environments \cite{dH23PhysFad}. 

The ultimate goal of our experiment is to demonstrate control over all of the scattering parameters of a given microwave billiard system. Often, the functional properties of a cavity are designed into the shape and structure of the cavity, usually including symmetries to help meet the design goals. In our experiments, by contrast, we use a chaotic cavity, which represents the most general wave scattering setting possible, containing no geometrical or hidden symmetries. To gain control over the system we instead place active metasurfaces inside the cavity, which effectively allow electronic manipulation of the boundary conditions of the cavity.  Scattering environments encountered in the real world are usually very complex and lack any symmetries. By demonstrating control over this complex and low-symmetry system, we are moving one step closer to actively controlling the scattering environments of arbitrary real-life systems.

\section{Experimental Setup}
\begin{figure}[htb]
  \subcaptionbox*{}[.45\linewidth]{%
    \includegraphics[width=6.8cm]{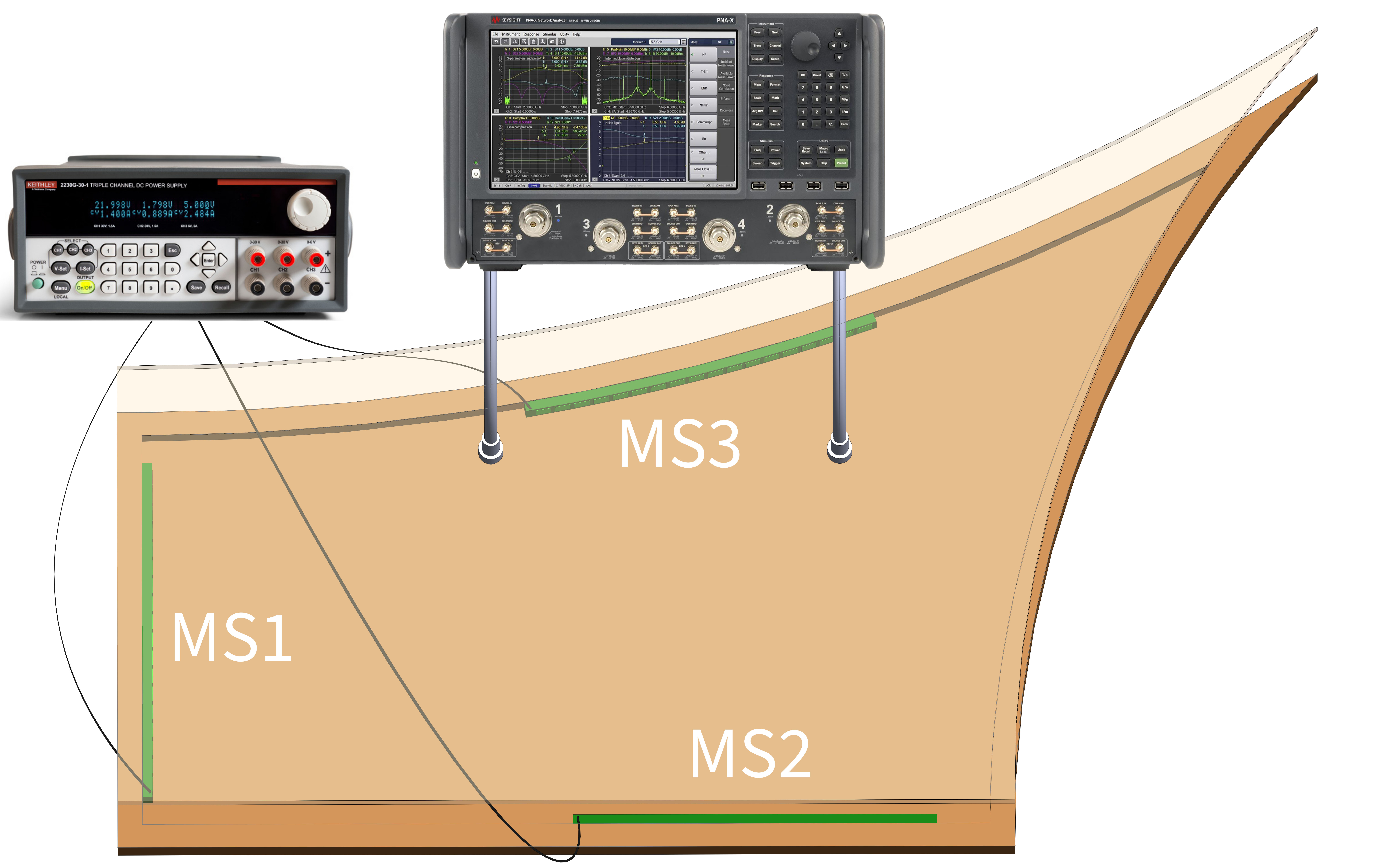}%
  }%
  \hspace*{\fill}
  \subcaptionbox*{}[.45\linewidth]{%
    \includegraphics[width=6cm]{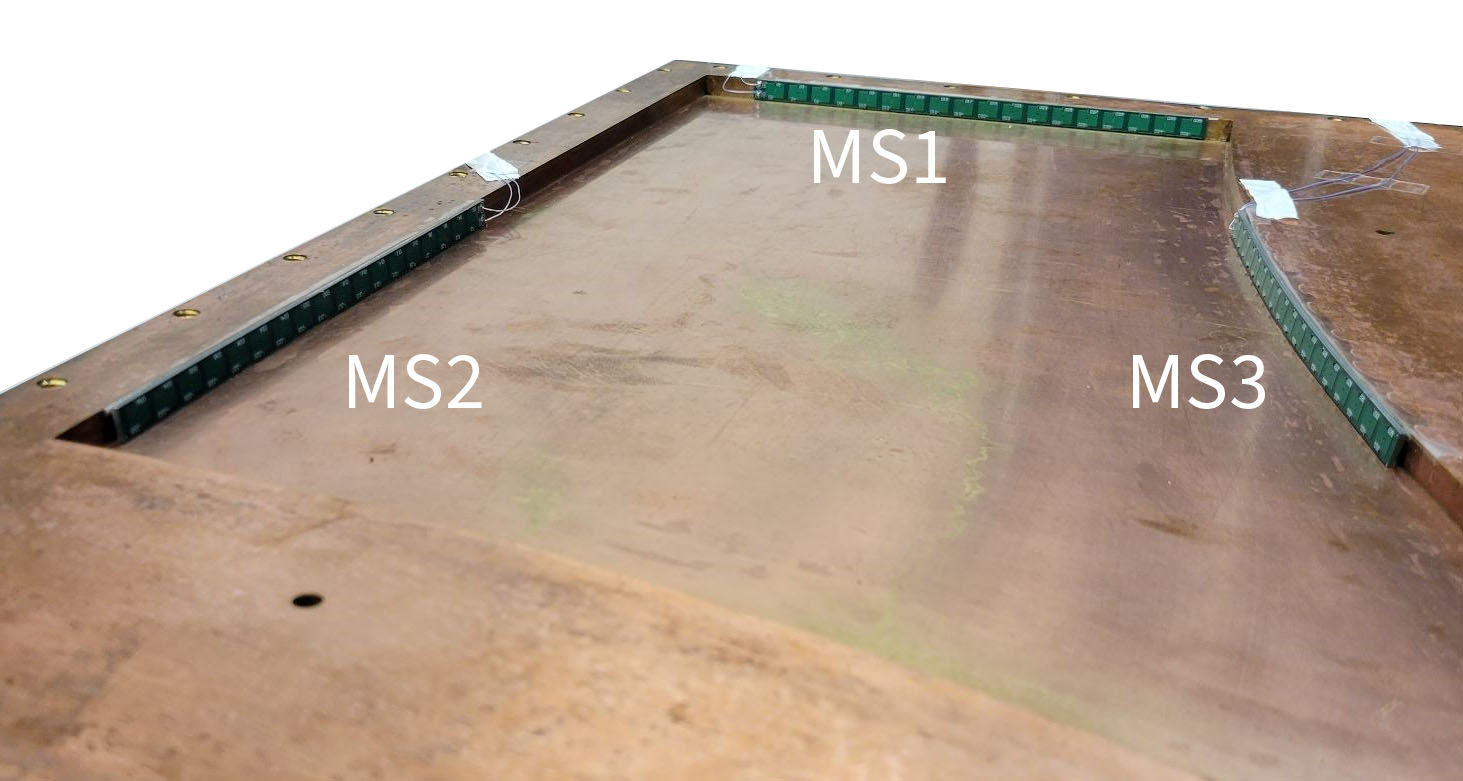}%
  }
  \caption{Quarter bow-tie billiard with three tunable metasurfaces (MS1-3) along the interior walls of the cavity. The left image is a schematic of the quarter bow-tie billiard with the three metasurfaces inside, and the lid lifted off the base. Two antennas are connected to the billiard through the lid and are attached to coaxial transmission lines, representing the scattering channels, and these are connected to the VNA. The right image is a perspective photograph of the interior of the base of the cavity, with the lid removed, showing the three metasurfaces.}
\label{Fig:1}
\end{figure}

A quasi-two-dimensional wave chaotic quarter bow-tie billiard is loaded with three tunable nonlinear metasurfaces. The billiard has a height of 7.9 mm and an area of 0.115 $m^2$. Therefore as long as the cavity is excited at frequencies below approximately 19 GHz, the system supports only one propagating mode, with electric field polarized in the vertical direction. The billiard has been used before to demonstrate the crossover from Gaussian Orthogonal to Gaussian Unitary ensemble statistics in both the level spacings \cite{So95} and eigenfunctions, \cite{Wu98,Chung00} and more recently has been loaded with microwave diodes to act as a reservoir computer \cite{Shukai22}. The metasurfaces used in this work were fabricated by the Johns Hopkins University Applied Physics Laboratory \cite{PhysRevApplied.20.014004} and were designed to be used for reflection amplitude variation between 11-18 GHz and reflection phase variation between 14-16 GHz. The metasurfaces are a linear array of 18 varactor-loaded mushroom-shaped resonant elements, where each element is sub-wavelength in size \cite{Siev99}. The PCB material of the metasurfaces is Rogers 5880 and the diode part number is MACOM MAVR-011020-1411. Each metasurface is 1.8 mm thick, 7.9 mm high, and 185 mm long, and is flexible enough to conformally attach to a curved interior wall.  The diodes on a given metasurface can be tuned simultaneously with a globally applied DC voltage bias to the metasurface through thin insulated wires that exit the cavity underneath the lid. As the applied voltage bias is increased from 0 to 15 V, the capacitance of the varactor diodes varies from 0.24 to 0.03 pF, thus increasing the resonant frequency of the patches. Note that as the voltage is varied, the reflection magnitude and phase of the metasurface both change, in general. Thus the tuned perturbation has a complex effect on the modes of the cavity. The metasurfaces are placed along three different walls of the billiard and are connected to a Keithley 2230G-30-1 triple channel programmable DC power supply, as shown in Fig. \ref{Fig:1}. Each of the three metasurfaces cover approximately 12\% of the perimeter of the billiard.  The quality factor of the billiard with the 3 metasurfaces present is approximately 250 at 9 GHz.

An Agilent Technologies model N5242A or Keysight model N5242B microwave vector network analyzer (VNA) is connected to the two ports of the billiard through coaxial cables that support a single mode of propagation, and the $2\times 2$ complex scattering matrix ($S$) is measured over a chosen frequency range. Through the variation of the reflection coefficient of the metasurfaces, we can manipulate the poles and zeros of the scattering matrix to create conditions for coherent perfect absorption (CPA), among other things.  CPA occurs when a zero of the scattering matrix is brought to the real frequency axis \cite{Chong10,LeiCPA20}.  When the system is excited with the $S$-matrix eigenvector that corresponds to the zero eigenvalue, all of the injected energy is absorbed inside the scattering system, and none is reflected or transmitted through any of the scattering channels. The perfect absorption condition requires this very specific type of coherent eigenvector excitation of the billiard, and any deviation causes significant reflection and/or transmission.

CPA requires that the system contain non-zero loss, either in a distributed or localized manner, or both. The waves sent into the system reverberate in such a way as to be completely absorbed, and in addition, so that no energy emerges from any of the ports of the system, either in transmission or reflection.  This condition is achieved by exciting all of the ports simultaneously at the (single) CPA frequency, but with appropriate relative amplitudes and phases on the incident waves on all of the ports.  With these precise coherent excitation conditions, one can achieve the above-stated outcomes. However, if any of the excitation conditions are modified, the CPA condition is lost, and this situation is explored experimentally in detail below. For an arbitrary scattering system it is very difficult to find the coherent excitation conditions for CPA by analytical or numerical means. We rely on measurements of the scattering matrix, and our ability to strongly modify the scattering properties of the cavity using the embedded metasurfaces, to establish the CPA conditions. In fact, our approach is so successful that CPA conditions can be established at almost any frequency in the bandwidth of operation of our experiment.

To establish CPA experimentally, we first choose a frequency range of interest and measure the scattering matrix with the VNA, and with each successive measurement the applied voltage bias of a particular metasurface is increased, usually with a step size of 0.01 V. During the set of measurements, the other metasurfaces in the cavity are held at a fixed applied voltage bias. From this set of measured $S$-matrices, the complex Wigner-Smith time delay \cite{PhysRevE.103.L050203} is calculated as a function of frequency at various applied biases to the metasurface. The conditions for CPA are found by identifying parametric settings where the magnitude of the complex time delay diverges.

Once the conditions for a CPA state are found, the scattering matrix is diagonalized to determine the $\lambda_S = 0+i0$ eigenvalue and the corresponding eigenvector $\ket{\psi_{CPA}}$. For our system, the scattering matrix has two eigenvalues $\lambda_S$ and eigenvectors $\ket{\psi}$ whose elements are two complex numbers representing the amplitudes of the waves injected at the two ports, but we can determine which one is the CPA excitation by finding the one that has an identically zero eigenvalue. From the CPA eigenvector, the relative amplitude and phase of the required excitation that must be simultaneously injected into the two ports is determined. With this information, the 2-port dual source mode of the VNA is used to inject the CPA eigenvector into the billiard to verify that the conditions for coherent perfect absorption are achieved. For $S$-parameter measurements, the VNA is calibrated up to the antennas on the billiard. But when the 2-port dual source mode is activated, the calibration is no longer valid. This causes the necessary parameters determining the CPA excitation to be slightly different than from the calibrated $S$-parameter measurement. During the CPA injection, we measure the ratio of outgoing power $P_{out, j}$ to incident power $P_{in,j}$ on each port $j$ as a function of several parameters near the CPA condition. To do this, the receivers on each port of the VNA are used to measure both the power that goes into the cavity and the power that comes back out.  

\section{Complex Wigner-Smith Time Delay}
The Wigner-Smith time delay $\tau_W$ is a rough measure of how long a wave lingers in a scattering system before leaving. In unitary systems, the time delay is a real quantity, but for subunitary systems it becomes complex valued \cite{Asano16,dH21CPA,PhysRevE.103.L050203}. The Wigner-Smith time delay is defined as: \begin{equation} \tau_{W} = - \frac{i}{M} \frac{\partial}{\partial f} log\:det\:S \end{equation} where $i$ is the imaginary number, $M$ is the number of ports connected to the system, $f$ is frequency, and $S$ is the scattering matrix. Experimentally, the frequency derivative in the Wigner-Smith time delay is calculated using a central difference formula.

It is known that the determinant of $S$ can be written as a Weierstrass factorization extending over the complex frequency plane as \cite{Sokolov89,Fyod97,FyodorovRMT11,Kuhl13,Grig13a,Grig13b,Schomerus17,Genevet23a,Genevet23b} 
\begin{equation}det S(f) \propto  \prod_{n = 1}^{N} \frac{f + i\eta -  z_n}{f + i\eta -  \varepsilon_n} \end{equation} where $N$ is the total number of modes of the closed system, $\eta$ is the uniform attenuation, and $z_n$ and $\varepsilon_n$ are the zeros and poles, respectively, of the scattering matrix. We further define $z_n$ and $\varepsilon_n$ as: \begin{equation}z_n =  Re[z_n] + i Im[z_n] \end{equation} \begin{equation}\varepsilon_n =  f_n -  i \Gamma_n, \end{equation} and adopt the convention that $\Gamma_n>0$ in passive lossy systems.  It has been shown \cite{PhysRevE.103.L050203} that the complex Wigner-Smith time delay associated with each mode $n$ can be written as a sum of two terms for both the real and imaginary parts, one involving the scattering poles and the other the scattering zeros, as follows. 

\begin{align}
\text{Re}\: \tau_{W} &= \frac{1}{M} \sum\limits_{n=1}^{N} \left[ \frac{Im [z_n] - \eta}{(f-Re [z_n])^2 + (Im [z_n] - \eta)^2} + \frac{\Gamma_n + \eta}{(f-f_n)^2 + (\Gamma_n + \eta)^2}\right]
\label{Retau}
\\
\text{Im}\: \tau_{W} &= -\frac{1}{M} \sum\limits_{n=1}^{N} \left[ \frac{f-Re [z_n]}{(f-Re[z_n])^2 + (Im [z_n] - \eta)^2} - \frac{f-f_n}{(f-f_n)^2 + (\Gamma_n + \eta)^2}\right]
\label{Imtau}
\end{align}

Using this formalism, we can see that the Wigner-Smith time delay is divergent at $f=Re[z_n]$ when the imaginary part of the zero $Im[z_n]$ of the scattering matrix is equal to the uniform attenuation of the system $\eta$. The diverging time delay for a wave propagating inside a lossy system is an indication that an $S$-matrix zero has crossed the real frequency axis, which is the enabling condition for coherent perfect absorption \cite{dH21CPA,PhysRevE.103.L050203}.   Further, from the measured Wigner-Smith time delay, the poles and zeros can be systematically extracted using Eqs. (\ref{Retau}) and (\ref{Imtau}), and their evolution with metasurface voltage bias can be visualized.  For example, note that the divergent term in the real part of $\tau_W$ has a sign that depends on the sign of $Im[z_n]-\eta$, which changes as the imaginary part of the $S$-matrix zero is varied.  Using information such as this, we can understand how a complex scattering system interacts with incoming waves, and use this knowledge to engineer conditions for achieving CPA.

\begin{figure}[htb]
\centerline{%
\includegraphics[width=8cm]{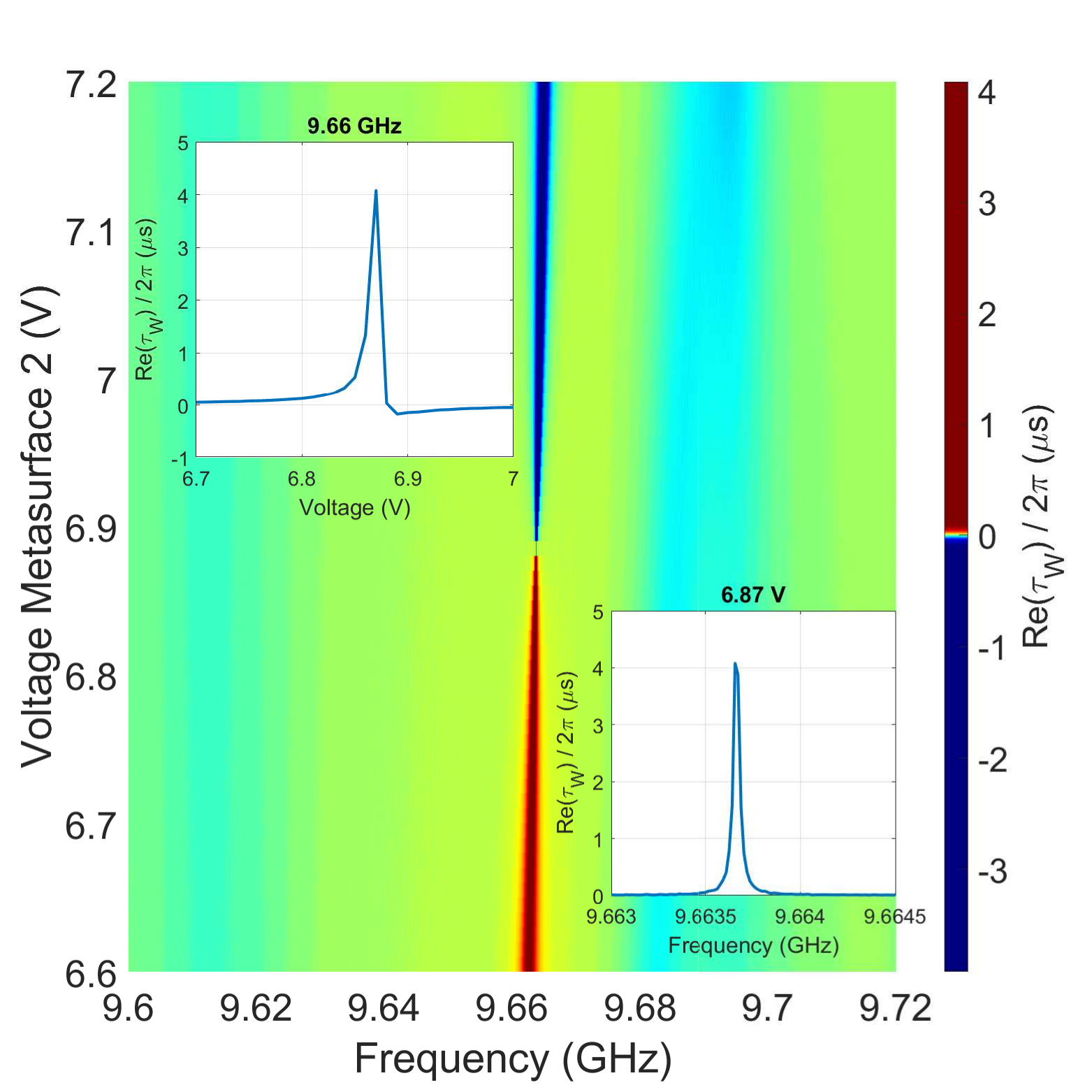}}
\caption{Real part of the Wigner-Smith time delay $Re[\tau_W]/2\pi$ vs. frequency and metasurface 2 voltage for a particular cavity realization. We use a nonlinear color scale to help distinguish the areas of small time delay from the areas of very large time delay due to the divergence of the Wigner-Smith time delay at one point in this parameter space.  With this color scale, the majority of the colors between blue and red (green-yellow) are concentrated near the mean value of the time delay in this parameter space ($\sim$3 ns).  The single largest value of the Wigner-Smith time delay measured in this domain has a value of 696.5 $\mu s$, but is excluded from this plot for clarity. The Heisenberg time ($\tau_H = \frac{2\pi}{\Delta}$, with mean mode-spacing $\Delta = \frac{c^2}{2\pi f A}$ for this 2D billiard of area $A$) at 9.66 GHz is $\tau_H \approx$ 0.49 $\mu$s.  Note that $\tau_W$ is divided by $2\pi$ to convert to seconds.  Left Inset: Real Wigner-Smith time delay vs. voltage at 9.66 GHz. Right Inset: Real part of Wigner-Smith time delay vs. frequency at a fixed voltage of 6.87 V.}
\label{Fig:2}
\end{figure}

\section{Experimental Results}
For a specific setup of the cavity, we take measurements over a particular frequency range and voltage bias applied to one of the metasurfaces. Then the applied bias is swept until the Wigner-Smith time delay shows a near divergence.   A typical result is shown in Fig. \ref{Fig:2} for the real part of the Wigner-Smith time delay in a narrow range of frequency and voltage. For the mode at 9.66 GHz, the time delay becomes very large in magnitude around a metasurface bias of 6.87 V. This indicates that a CPA state is possibly located at the point of divergence.  The left inset of Fig. \ref{Fig:2} shows the real part of the time delay  as the voltage bias is increased at a fixed frequency of 9.66 GHz, through the point of divergence.  The dramatic increase can be explained by the imaginary part of scattering zero ($Im[z_n]$) decreasing in magnitude toward the value of the uniform attenuation ($\eta$) of the system, causing the Wigner-Smith time delay to increase. At the divergence, $Im[z_n]$ is equal to $\eta$, and past the point of divergence, $Im[z_n]$ has decreased below the value of $\eta$, causing the sign of the Wigner-Smith time delay to switch from positive to negative (see Eq. (\ref{Retau})). In the right inset of Fig. \ref{Fig:2}, the voltage is kept fixed at 6.87 V while the frequency is increased through the point of divergence.  A similar increase of the real part of the time delay is observed, except that the sign of the Wigner-Smith time delay remains positive through the divergence. The interpretation is that at a fixed voltage bias of 6.87 V, the values of $Im[z_n]$ and $\eta$ remain essentially fixed as the frequency is varied, while the divergence arises from the frequency dependence indicated in Eq. (\ref{Retau}).  The variation of $Re[\tau_W]$ from +4 $\mu s$ to -3.6 $\mu s$ visible in Fig. \ref{Fig:2} occurs diagonally across frequency and voltage bias because the bias voltage changes both the real and imaginary parts of the S-matrix zero simultaneously. 

\begin{figure}[htb]
\centerline{%
\includegraphics[width=0.80\textwidth]{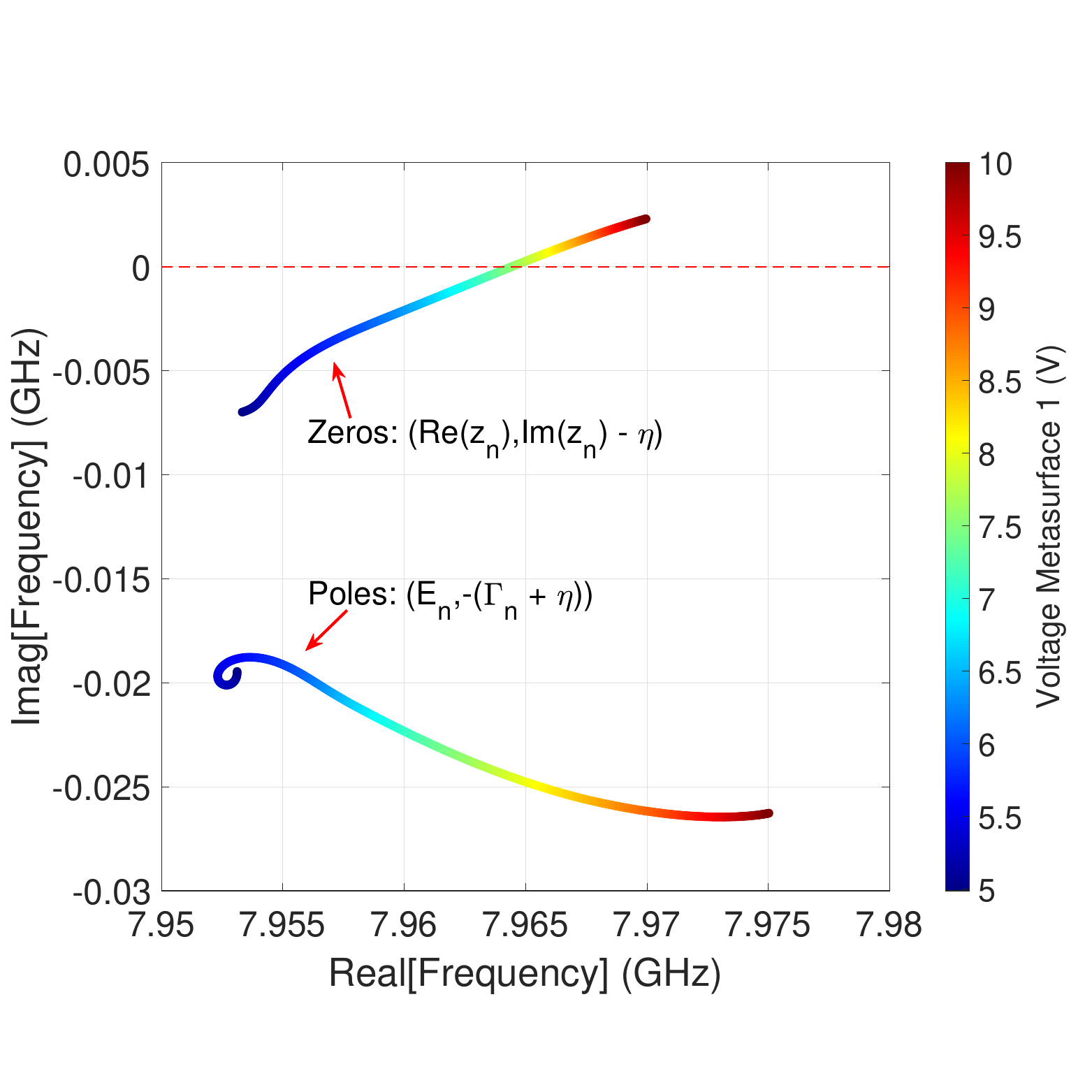}}
\caption{Zeros and poles of the scattering matrix for a particular mode of the bowtie billiard, extracted from measured complex time delay, plotted in the complex frequency plane as a function of voltage bias applied to metasurface 1. For each successive point from blue to red, the applied bias to metasurface 1 is increased while holding all other parameters fixed. The top curve shows evolution of the $S$-matrix zero, and the bottom curve shows evolution of the pole.}
\label{Fig:3}
\end{figure}

Using the Wigner-Smith time delay expression as a sum of two terms for the real and imaginary parts, the location of the poles and zeros of the scattering matrix are extracted as functions of applied bias to metasurface 1 for another mode of the bowtie billiard, as shown in Fig. \ref{Fig:3}. This is accomplished by fitting the Lorentzian expressions (Eqs. (\ref{Retau}) and (\ref{Imtau}) for a single mode) to the real and imaginary parts of the experimental data simultaneously, and from the best fitting we extract the pole and zero information of that single mode for each voltage bias.  Note that the fit parameters are $Re[z_n]$ and $Im[z_n]-\eta$ for the zero, and $f_n$ and $\Gamma_n +\eta$ for the pole, for each mode $n$. Figure \ref{Fig:3} shows that as the applied bias is increased, the zeros are observed to move upward in the complex frequency plane toward the real frequency axis, and CPA is enabled at the point where $Im[z_n]-\eta=0$. With this information, we now know how to engineer the cavity to have a specific time delay, degree of absorption, etc., for this frequency range.

\begin{figure}[htb]
\centerline{%
\includegraphics[width=0.75\textwidth]{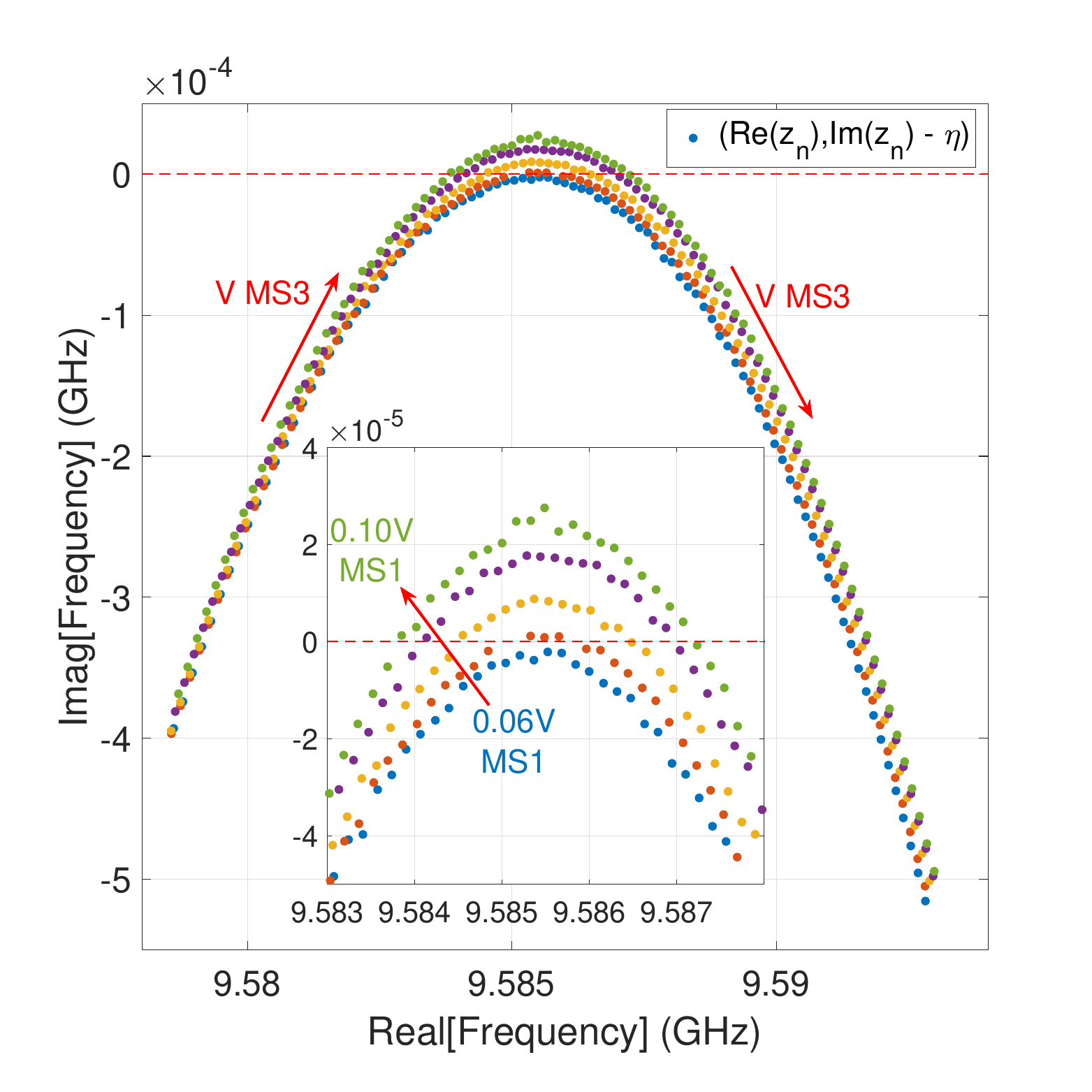}}
\caption{Zeros of the scattering matrix ($Re[z_n]$,$Im[z_n]-\eta$) for a particular mode of the bowtie billiard, plotted in the complex frequency plane. Each successive point from left to right is a 0.01V step increase in the voltage applied to metasurface 3 (MS3), and each color represents a different fixed voltage of metasurface 1 (MS1). During these measurements, metasurface 2 was held at a fixed voltage bias. The inset shows a zoomed in view of the real frequency axis near the axis crossings.}
\label{Fig:4}
\end{figure}

Taking measurements of the cavity under different conditions, a particularly interesting case is found where a single mode has a zero that crosses the real axis twice as the voltage bias to metasurface 3 is varied (see Fig. \ref{Fig:4}). These real frequency-axis crossings for the $S$-matrix zero correspond to two CPA states in the voltage sweep of one metasurface. Then, using the other two metasurfaces in the cavity, it is possible to raise and lower this parabolic evolution of the zeros in the complex plane, as illustrated in the inset of Fig. \ref{Fig:4}.  This result  demonstrates control of the scattering zeros of the system through the additional degrees of freedom that the other metasurfaces offer.

\begin{figure}[htb]
\centerline{%
\includegraphics[width=0.75\textwidth]{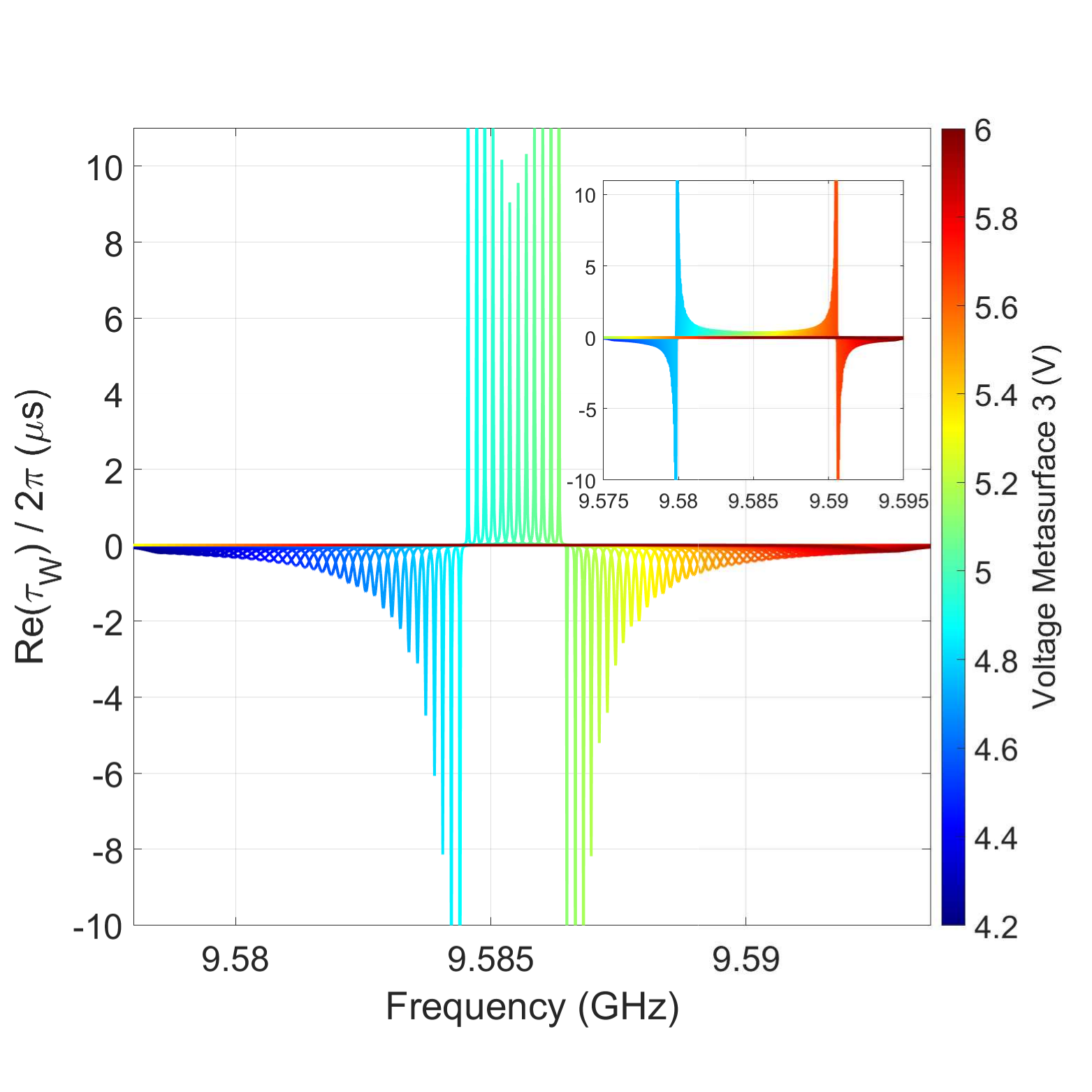}}
\caption{Real part of the Wigner-Smith time delay $Re[\tau_W]/2\pi$ vs. frequency recovered from a particular curve from Fig. \ref{Fig:4} describing the S-matrix zero evolution, and evaluated as a function of frequency using Eq. \ref{Retau}. The sign changes of the extreme values of time delay bracket the two CPA points. The lower frequency CPA state has an extreme time delay value around 60 $\mu s$, and the upper frequency CPA state has an extreme time delay value near -500 $\mu s$, not shown for presentation clarity.  The gaps in between the peaks for each color result from the finite metasurface bias voltage resolution of the measurement. The inset shows the real part of Wigner-Smith time delay vs. frequency for different fixed values of the metasurface 1 and 2 voltage biases, showing the two CPA points separated by a much larger frequency.}
\label{Fig:5}
\end{figure}

Using the extracted pole and zero locations as a function of voltage, we can use the Lorentzian expression (Eq. (\ref{Retau})) to examine in detail the real part of the Wigner-Smith time delay for one of the curves from Fig. \ref{Fig:4}.  In Fig. \ref{Fig:5} we see that there are two locations where the time delay is nearly divergent, and that in between the divergences there is a significant peak value of time delay, approximately 9 $\mu$s, and its location can be finely tuned in frequency. This delay corresponds to 2.7 km of free space travel, and with the characteristic length scale of the cavity being $\sqrt{A}$ (with cavity area $A=0.115\ m^2$), this corresponds to approximately 8,000 bounces around the cavity.  It should be noted that at all bias values the time delay-bandwidth product is on the order of unity. From Fig. \ref{Fig:5}, we can also see from the inset that we can move the scattering matrix zeros and push apart the frequencies of diverging time delay by utilizing the other metasurfaces inside the cavity. This also causes the peak time delay between the two zero crossings to decrease.

\color{black}

Once the conditions where the Wigner-Smith time delay diverges are found, we see that the $S$-matrix has at least one eigenvalue nearly equal to $\lambda_S = 0+i 0$. We can calculate the eigenvector of the $S$-matrix corresponding to this eigenvalue.  To test for the CPA condition, we must inject this eigenvector excitation into the billiard.   The CPA eigenvector can be defined as: $\ket{\psi_{CPA}} = \begin{psmallmatrix} X e^{i\theta} \\ Y e^{i\psi} \end{psmallmatrix}$, defining the amplitudes and phases of the excitation at the CPA frequency. Hence for the CPA excitation, the relative phase of the signal between ports 1 and 2 is $\psi-\theta$, and the relative amplitude between ports 1 and 2 is $20\log_{10} \left(\frac{X}{Y}\right)$, when $X$ and $Y$ are measured in linear voltage and need to be converted to dB. The VNA is used to inject this specific excitation into the system at the CPA frequency. The expectation is that the output vector $\ket{\psi_{out}} = S \ket{\psi_{in}} = 0$ in this case, meaning that all of the input energy is perfectly absorbed, and none is reflected or transmitted. 

\begin{figure}[htb]
\centerline{%
\includegraphics[width=\textwidth]{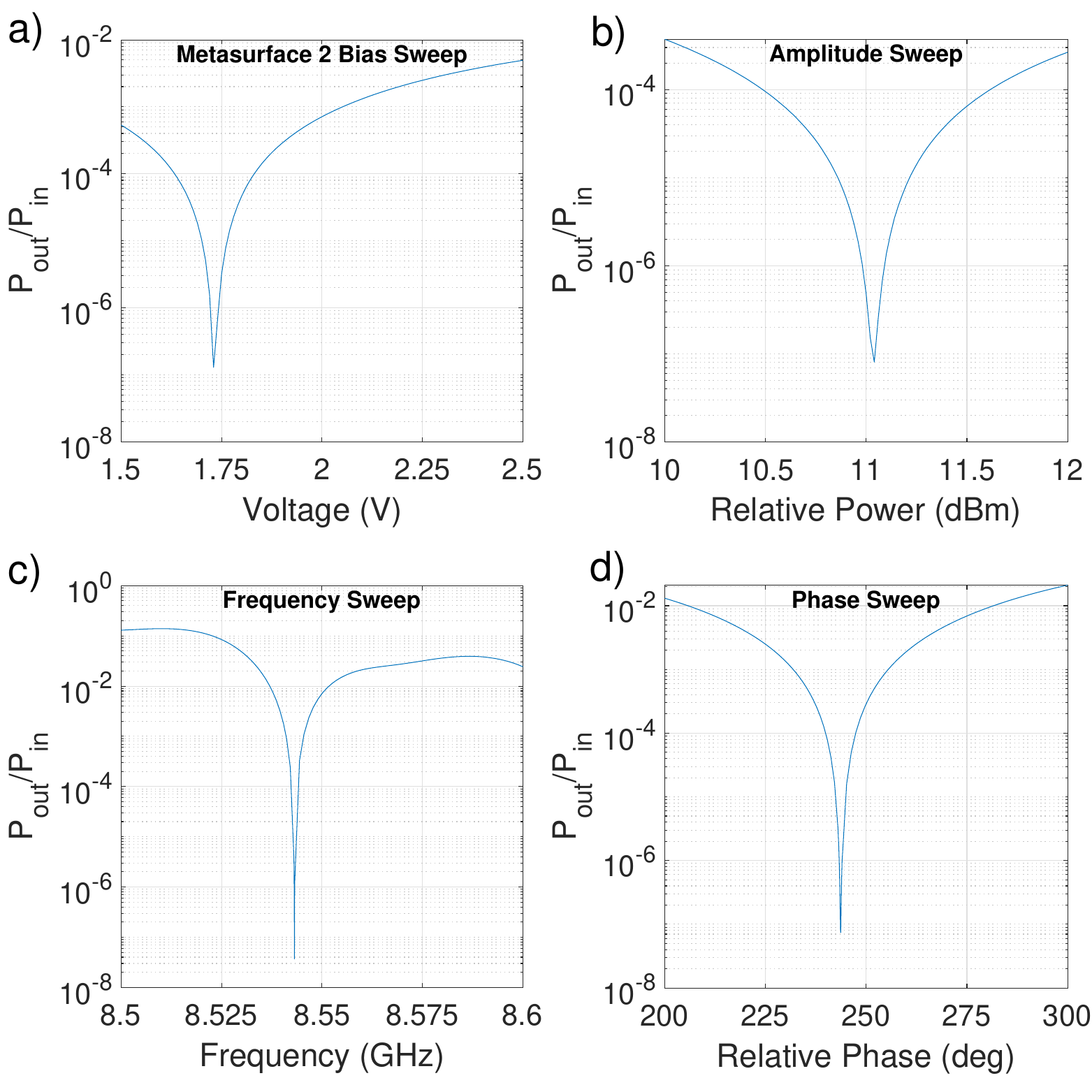}}
\caption{Evidence of Coherent Perfect Absorption in the ray-chaotic quarter bow-tie microwave billiard, for four independent parametric sweeps.  We measure the ratio of the total output power to the total input power, $P_{out}/P_{in}$ as a function of parameter variation. The swept voltage bias to metasurface 2 is shown in (a). Small parametric deviations from the CPA input eigenvector $\ket{\psi_{CPA}}$ are made through the relative power $20\log_{10} \left(\frac{X}{Y}\right)$ (b), frequency (c), and relative phase $\psi-\theta$ (d) sweeps.}
\label{Fig:6}
\end{figure}

To test this experimentally, we measure both the injected and received powers from the cavity, and see how those powers vary as the parameters of the system, and the excitation, are changed. In Fig. \ref{Fig:6}, the four parameters we have control over are systematically varied to verify that the CPA state is correctly identified. The four parameters are the voltage applied to the metasurface (which controls the location of the S-matrix zero in the complex frequency plane), the relative amplitudes of the signals on the two input channels, the frequency $f$, and the relative phase of the signals injected into the two channels. For each parameter sweep, the other parameters are set to their optimal values. We measure the total input power delivered to the system through the two channels, $P_{in}=P_{in,1}+P_{in,2}$, and the total power that emerges from the cavity, $P_{out}=P_{out,1}+P_{out,2}$, and form the ratio $P_{out}/P_{in}$.  This process is carried out in an iterative manner and converges to a point very close to the CPA condition for the system, which corresponds to $P_{out}/P_{in}=0$.  For example, in the frequency sweep in Fig. \ref{Fig:6}(c), the ratio of output power to input power changes by approximately seven orders of magnitude for small changes in frequency, with a minimum output to input power ratio of $\frac{P_{out}}{P_{in}} = 3.71 \times 10^{-8}$.  We observe that the exact conditions of the CPA state slowly drift over time, likely due to the temperature fluctuations of the laboratory. We believe this temperature drift to be the limiting factor on how precisely we can measure the CPA state. This illustrates how sensitive the CPA state is to small perturbations to the system, making it attractive for use as a sensor. 

\section{Discussion}
Much of the prior work to controllably alter the scattering matrix of complex systems has been done by mechanical means, mainly for the purpose of perturbing the poles of the scattering matrix \cite{PhysRevE.73.035201}. With the metasurfaces, the poles and zeros of the scattering matrix can be smoothly varied in the complex frequency plane and placed in determined locations, purely by electronic means. By moving the zero of the scattering matrix to the real frequency axis, we can enable CPA and this is correspondingly seen as a divergence in the real part of the Wigner-Smith time delay. We can inject the zero-eigenvalue eigenvector and observe nearly complete absorption, but this is a singular condition which is only possible at a single point in the parameter space. Suwunnarat \textit{et al.} have recently shown the creation of a nonlinear CPA by using exceptional point degeneracies of the zeros of the scattering matrix \cite{Suwunnarat_2022}. Work such as this shows the future possibilities of creating larger parametric regions of coherent near-perfect absorption, which can be beneficial for a variety of applications in communications and wireless power transfer.
 
\section{Conclusions}
In this paper we show that using tunable metasurfaces inside a two-dimensional wave-chaotic cavity, we can control the locations of scattering poles and zeros in the complex frequency plane. By perturbing the system with the metasurfaces, we can drag the scattering zeros across the real frequency axis to create a coherent perfect absorption condition. The demonstration of nearly complete absorption is illustrated in Figure \ref{Fig:6} by having approximately seven orders of magnitude less power leaving the system compared to that injected in the CPA eigenvector. We also demonstrated precise manipulation of the location of the CPA state by applying voltage biases to the other metasurfaces in the cavity. This is equivalent to controlling the scattering properties of the system, and this control gives us the ability to engineer specific conditions in the cavity, such as having coherent perfect absorption at a particular frequency, tuning a desired time delay for a signal, having regions of high absorption for unwanted signals, and low absorption for desired signals, etc.

\section*{Acknowledgements}
We acknowledge insightful discussions with Tsampikos Kottos, Dan Sievenpiper, and Lei Chen.  This work was supported by NSF/RINGS under grant No. ECCS-2148318, ONR under grant N000142312507, ONR under grant N0002413D6400, DARPA WARDEN under grant HR00112120021, ONR DURIP FY21 under grant N000142112924, and ONR DURIP FY22 under grant N000142212263.

\bibliographystyle{unsrt}
\bibliography{Bibliography}

\end{document}